\documentclass[12pt]{amsart}
\usepackage{amsmath,amsfonts,amssymb,amscd,amsthm}
\usepackage{graphics}
\usepackage[frenchb]{babel}

\theoremstyle{definiction}

\theoremstyle{remark}

\newcommand{\mq}{M\'ecanique Quantique}
\newcommand{\mc}{M\'ecanique Classique}
\newcommand{\s}{Schr\"odinger}
\newcommand{\id}{ind\'eterminisme}
\newcommand{\ip}{impr\'edictibilit\'e}
\newcommand{\et}{\'etats coh\'erents}
\newcommand{\cad}{c'est-\`a-dire }
\newcommand{\sci}{sensibilit\'e aux conditions initiales}
\newcommand{\lsc}{limite semiclassique}

\title{Ind\'eterminisme quantique et impr\'edictibilit\'e classique}
\author[T. Paul]{Thierry Paul}
\address{centre de Math\'ematiques laurent Schwartz, Ecole polytechnique et CNRS}
\email{paul@math.polytechnique.fr}
\begin{document}
\date{}
\maketitle
\tableofcontents
\section{Introduction}
On a souvent fait le reproche \`a la \mq\  de manquer  la plus belle des  propri\'et\'es de la \mc\ : le d\'eterminisme. 

``... il faut r\'esoudre les \'equations..." \cite{VA} semble \^etre le credo classique, coupl\'e \`a l'id\'ee que la 
solution de telles \'equations donne une description compl\`ete des ph\'enom\`enes. N\'ee justement afin de palier
une incompl\'etude de la \mc, celle qui fait que deux ``points mat\'eriels" attir\'es l'un vers l'autre par
la loi de la gravitation universelle se choquent au bout d'un temps fini et sortent alors du paradigme classique, 
 la \mq\ a, d\`es ses d\'ebuts, n\'ecessit\'e l'introduction de l'al\'ea : la dynamique quantique est certes stable, mais vit dans un ``espace" qui ``nous"
 est inaccessible dans sa totalit\'e, et  l'acc\`es au syst\`eme quantique se fait une projection sur
 un bord de cet espace, au moyen du  processus de mesure. De plus cette projection est al\'eatoire, comme nous le pr\'eciserons plus bas.
 
 Faire un proc\`es \`a la \mq, plut\^ot \`a son ind\'etermisnime, c'est oublier un peu vite le traumatisme cr\'e\'e
 par Poincar\'e \`a la fin du XIX\`eme si\`ecle lors de son abandon du paradigme laplacien de stabilit\'e
 \cite{PO}. De petites variations des conditions initiales (\`a l'origine des petites
 variations des param\`etres de l'\'equation de la dynamique) font appara\^itre de grands effets. Le   joyeux d\'eterminisme 
 laisse la place, en particulier dans la pratique, \`a une \ip\ soucieuse. Pour de tr\`es longs temps d'\'evolution
 cette \ip\ prend le dessus sur le d\'eterminisme et par id\'ealisation du passage \`a la limite des temps
 infinis, l'\ip\ se change carr\'ement en \id.
 
 Ces deux al\'eatoires sont de nature fort diverse : le classique \'etant extrins\`eque, d\`u \`a notre manque d'information
 totale sur les conditions initiales, alors que l'al\'eatoire quantique est intrins\`eque, la violation des in\'egalit\'es de Bell en a donn\'e la preuve.
  
  Nous allons tenter dans cet article de pr\'esenter de fa\c con informelle des r\'esultats r\'ecents qui mettent en jeu un possible recouvrement entre ces deux notions d'al\'eatoire.
\section{La \mq}
L'ind\'eterminisme quantique est un ind\'eterminisme d'acc\`es. Il surgit lorsque nous essayons de conna\^\i tre le syst\`eme, d'\'etablir certaines de ses propri\'et\'es, bref
d'effectuer une ``mesure". L'\'evolution \textit{strictement} quantique est tout \`a fait d\'eterministe : 
que ce soit dans le cadre de la m\'ecanique  des matrices ou de celui de l'\'equation de 
\s, le syst\`eme \'evolue en suivant une ``\'equation", aussi belle que celles de la \mc. Mais peut-on parler raisonnablement
d'\'evolution strictement quantique, c'est \`a dire de l'\'evolution
d'un syst\`eme sans notre d\'esir de le conna\^\i tre? Comment m\^eme pourrait-on imaginer, dans la science moderne, exempte de tout obscurantisme, une telle chose? D'autres s'en sont
charg\'es, la science se doit de se confronter \`a l'observation. C'est un point qu'elle partage, d'ailleurs, avec la philosophie.

Il n'y a donc qu'\textit{une} \'evolution quantique, et elle contient pleinement le ph\'enom\`ene 
de la mesure. Et c'est l\`a que l'\id\ intervient. Moyen \'economique d'acc\`es au ``monde"
quantique, extraordinairement puissant et coh\'erent.

Nous allons passer en revue  les quatre axiomes de la \mq, sans rentrer dans les d\'etails 
(voir \cite{p3} pour un expos\'e plus pr\'ecis, quoique non technique).

Un syst\`eme quantique est d\'ecrit par un vecteur, et donc l'ensemble des \'etats d'un syst\`eme est un espace vectoriel. 
C'est l'{\bf Axiome I}\rm, qui fixe le cadre
cin\'ematique et dont il faut retenir surtout le fameux principe de superposition, l'id\'ee qu'il y a un $+$, et que l'on peut sommer des \'etats. La juxtaposition, et surtout
l'interaction de deux (sous-)syst\`emes met en jeu la notion math\'ematique de produit tensoriel ; l'{\bf Axiome II}\rm\ 
nous dit comment ``fabriquer" l'espace des \'etats : on
prend le produit tensoriel des deux espaces d'\'etats des deux (sous-)syst\`emes. La notion de produit tensoriel n'est pas simple, et repr\'esente une extension de la notion
habituelle de produit : sachant que l'espace des \'etats doit \^etre un espace vectoriel, il faut consid\'erer non seulement les produits d'\'etats originaux provenant des deux 
(sous-)syst\`emes, mais aussi les sommes de tous ceux-ci \footnote{pour \^etre rigoureux toutes les combinaison lin\'eaires}. Et c'est bien l\`a que la saveur quantique prend
toute sa dimension : la somme de deux \'etats produits n'est plus forc\'ement un \'etat produit. 
C'est dans ce ph\'enom\`ene d'intrication que la nature quantique  \'echappe \`a la culture classique,
et c'est ici que la mesure sera n\'ecessaire. Une premi\`ere partie de l'\'evolution est donn\'ee par l'{\bf Axiome III}\rm\ qui stipule une \'evolution unitaire, c'est \`a dire
qui pr\'esente cette id\'ee tellement ch\`ere \`a la science moderne : la conservation d'une quantit\'e. 
Ici la quantit\'e conserv\'ee sera la norme du vecteur-\'etat, que l'axiome suivant
va interpr\'eter en termes de somme (totale) de probabilit\'es.

En effet l'{\bf Axiome IV}\rm\ permet, entre autres, de palier les effets d\'evastateurs (d\'evastateurs de notre culture 
classique, mais non de notre id\'ee de coh\'erence) de l'axiome
II. Effectuer une mesure c'est projeter l'\'etat du syst\`eme sur un vecteur particulier, correspondant \`a un r\'esultat possible de la mesure. Cette valeur particuli\`rere (par
exemple la position), et donc le vecteur correspondant, sont ``choisis" au hasard, hasard intrins\`eque \`a la th\'eorie.

Nous renvoyons le lecteur \`a \cite{p3} pour plus de d\'etails et de g\'en\'eralit\'e, mais, pour ce qui va nous int\'eresser ici, nos allons nous concentrer sur la probabilit\'e de pr\'esence
associ\'ee \`a la fonction d'onde. Un syst\`eme quantique qui ne poss\`ede que des degr\'es de libert\'e 
classiques, c'est-\`a-dire position-impulsion, est repr\'esent\'e par une
fonction d'onde, d\'efinie sur l'espace et \`a valeur complexe. Lorsque l'on dit que le module au carr\'e de la
fonction d'onde repr\'esente une probabilit\'e de pr\'esence, on
entend que, lorsque l'on effectue la mesure correspondant \`a la position, l'{\bf Axiome IV}\rm\ stipule que le r\'esultat peut \^etre un point quelconque de l'espace, et que si
l'on effectue plusieurs fois la mesure sur un syst\`eme dans le m\^eme \'etat $\psi$, on obtiendra la valeur $x$ avec une probabilit\'e $|\psi(x)|^2$. Bien que cette formulation 
fasse r\'ef\'erence explicitement \`a une collection d'\'evenements, il faut remarquer que les postulats de la \mq\ 
envisage des \'ev\`enements uniques, et que l'aspect probabiliste
r\'ev\`ele toute sa saveur dans les cas extr\^emes : le cas o\`u la probabilit\'e est nulle et celle o\`u elle prend 
la valeur maximale, c'est-\`a-dire $1$. Ce dernier cas est
obtenu, et se doit de l'\^etre sous peine de manque de continuit\'e, lorsque l'on effectue une seconde mesure ``juste apr\`es" la premi\`ere. Cette propri\'et\'e est valid\'ee par la
deuxi\`eme partie de l'{\bf Axiome IV}\rm\ qui dit que, lors de la mesure, le vecteur d'\'etat est projet\'e sur un vecteur particulier associ\'e au r\'esultat de la mesure : celui qui pr\'edit avec probabilit\'e $1$ la m\^eme valeur du r\'esultat lors d'une mesure post\'erieure. 
Ainsi on se trouve amen\'e \`a une repr\'esentation de la \mq\ dans
laquelle une mesure est associ\'ee \`a un op\'erateur (matrice) dont les valeurs propres sont 
les r\'esultats possibles et les vecteurs propres associ\'es sont ces fameux
vecteurs d'\'etats {\it post-mesure}\rm. Et il suffit, pour que l'axiomatique soit ``coh\'erente", que la probabilit\'e d'obtenir la valeur $x$ soit le module au carr\'e du
produit scalaire de l'\'etat du syst\`eme avec le vecteur propre correspondant.

Il est d'usage de pr\'esenter la \mq\ comme une modification, une d\'eformation de la \mc. Mais 
la \mq\ \'etant, jusqu'\`a pr\'esent, la th\'eorie fondamentale de la physique, il
conviendrait plut\^ot de prendre l'habitude de voir la \mc\ comme une situation limite, 
un bord de la th\'eorie quantique. Ceci peut \^etre r\'ealis\'e simplement gr\^ace \`a l'usage des
\et. 

Introduits d\`es la s\'erie des articles fondateurs de Schr\"odinger en 1926, les \et\ constituent 
une famille de vecteurs d'\'etats, d\'ependant de la constante de Planck 
$\hbar$ et indic\'es par deux variables : une de position $q$ et une d'impulsiion $p$. On peut les voir comme la famille d'\'etats minimisant les in\'egalit\'e de Heisenberg
\begin{equation}\label{heis}
\Delta P\Delta Q\geq \frac\hbar 2
,\end{equation}
 mais
les deux propri\'et\'es fondamentales dont ils jouissent sont, tout d'abord, qu'ils g\'en\`erent tout l'espace 
des \'etats, \`a la fa\c con d'une base (non-orthonorm\'ee)
 ``continue", c'est-\`a-dire que tout vecteur d'\'etat se d\'ecompose sur les \et\ ; 
 ensuite que la famille des \et\ reste presque invariante par l'\'evolution quantique
 g\'en\'er\'ee par l'\'equation de \s. Le ``presque" \'etant guid\'e par la constante de Planck, et devenant exact lorsque celle-ci s'annule. Ainsi donc, \`a la limite
 semi-classique o\`u $\hbar\to 0$, l'\'evolution quantique, lorsque la  condition initiale est un de ces \et , se confond avec un flot sur cette famille, cette sous-vari\'et\'e de l'espace des \'etats.
 C'est ainsi que ``l'espace (de phase) classique", espace des positions-impulsions, se (re)trouve d\'ej\`a pleinement dans la \mq, comme espace des indices d'une famille de
 vecteurs particuliers \'evoluant ``entre-eux", si l'on peut dire.
 
 Quelques remarques s'imposent. Tout d'abord faut-il vraiment que $\hbar\to 0$? C'est-\`a-dire 
 doit-on changer la th\'eorie pour que la \mc\ appara\^isse? La r\'eponse est non : la
 \mc\ est vraiment dans la \mq, au bord de celle-ci. Il suffit pour s'en convaincre de remarquer que toutes les propri\'et\'es \`a $\hbar\to 0$ se retrouvent par exemple dans
 l'atome d'hydrog\`ene lorsque l'on consid\`ere les \'etats tr\`es excit\'es proche du continuum d'\'energie. Ensuite ce r\'esultat d'invariance de la famille d'\et\ par
 l'\'evolution quantique est-il valable \`a tout temps? Quid de l'impossibilit\'e d'associer une trajectoire 
 (voir \cite{bal} dans ce m\^eme volume)? La r\'eponse \`a la
 premi\`ere question est bien s\^ur n\'egative est sera le c\oe ur de cet article. En \'evoluant un \'etat coh\'erent se d\'eforme et plus le temps augmente plus cette d\'eformation est
 significative. A la limite des temps longs elle devient m\^eme g\^enante, nous en parlerons plus bas. 
 Retenons seulement pour l'instant que l'espace de phases de la \mc\ 
 appara\^it dans la \mq\ et que l'on peut ainsi d\'efinir l'\'evolution classique en suivant l'\'evolution 
 quantique des \et.
 
 \section{La \mc}
 Une diff\'erence fondamentale entre quantique et classique r\'eside dans 
 l'id\'ee d'un espace absolu pour la \mc. L'\'evolution classique se d\'eroule 
 en effet dans un espace
 fixe, d'acc\`es imm\'ediat pour l'observateur, et qui n'est pour ainsi dire que 
 le r\'eceptacle de la th\'eorie \cite{car}. Nous avons (entre autres) vu 
 dans la section pr\'ec\'edente comment cet
 espace apparaissait, indi\c cant les \et, famille sp\'eciale de vecteurs 
 d'\'etats. Comment appara\^\i t le cot\'e ``absolu" d'un tel espace? 
 Et comment appara\^\i t-il dans une
 th\'eorie hautement non-absolue, puisque baignant dans l'al\'eatoire suite \`a 
 l'{\bf Axiome IV}\rm ? C'est gr\^ace aux probabilit\'es justement, 
 plus exactement \`a leurs extr\^emes.
 
 Les in\'egalit\'es de Heisenberg (\ref{heis}) qui sont, rappelons le, un Th\'eor\`eme et non 
 un principe, stipulent que l'on ne peut pas \^etre mieux localis\'e autour d'un
 point de l'espace de phases qu'\`a une \'echelle $\hbar$. Mais justement, rien
 n'empeche de consid\'erer des familles d'\'etats, d\'ependant de $\hbar$, et tels 
 que $\Delta P\Delta Q\sim \frac\hbar 2$. De tels \'etats peuvent avoir
  la propri\'et\'e de ``tendre" vers un point de l'espace de phases, i.e. lorsque
  $\Delta P\sim\Delta Q\sim\sqrt{\frac \hbar 2}$. Mais nous verrons plus loin que ce
  n'est pas la seule possibilit\'e, et que des \'etats satisfaisant 
  $\Delta P\Delta Q\sim \frac\hbar 2$ avec $\Delta P\sim\hbar/2$ et 
  $\Delta Q\sim 1$ (ou m\^eme $\Delta Q\sim \frac 1 {\sqrt\hbar}$) peuvent \^etre aussi int\'eressants et apparaissent aussi. Mais
  revenons au cas $\Delta P\sim\Delta Q\sim\sqrt{\frac \hbar 2}$.
  
  Dans la limite $\hbar\to 0$, l'aspect probabiliste de la th\'eorie quantique ne
  dispara\^\i t pas, mais plut\^ot se singularise en n'autorisant plus que $0$ et $1$
  comme valeur possible de la loi de probabilit\'e. C'est l'aspect absolu qui entre
  en jeu. Une fois cette remarque faite,  il ne reste plus  qu'a montrer que le ``support" d'un \'etat
  coh\'erent \'evolue suivant les lois de la dynamique classique pour avoir une image
  compl\`ete.
  
  C'est donc un premier exemple de confrontation entre ind\'etermisnime quantique et,
  cette fois, pr\'edictibilt\'e classique ; le point mat\'eriel.
  Nous verrons plus loin comment l'hypoth\`ese  $\Delta P\sim\hbar/2\  
  \Delta Q\sim 1$ ruine cette vision, mais nous avons besoin tout de suite de
  regarder d'un plus pr\`es le sacro-saint d\'eterminisme classique.
 
 Finissons cette section en insistant encore sur  un malentendu tr\`es courant : comment la
 constante de Planck peut-elle tendre vers z\'ero? Comme nous l'avons d\'ej\`a dit, la limite semi-classique n'est 
 pas en
 dehors de la th\'eorie quantique, elle en est un bord, rappelons-nous l'exemple de  l'hydrog\`ene.
 
 %Enfin il convient, selon moi, de remarquer l'aspect ``incomplet" de la th\'eorie
 %quantique, justement \`a cause du boulet que constitue le `point mat\'eriel".
 
% Newton et le choc.
 
 %stabilit\'e gr\^ace a la non commutativit\'e
 
 %longue distance et relativit\'e/
 
 \section{La \sci}\label{sci}
 La m\'ecanique classique est, quant \`a elle, parfaitement d\'eterministe.  Cependant le
 ph\'enom\`ene de la \sci\ a chang\'e, \`a la fin du XIX\`eme si\`ecle et gr\^ace
 aux travaux de Poincar\'e, la vison de la notion de stabilit\'e. La \sci\ 
 exprime que, quelle que soit la proximit\'e de deux conditions initiales, on
 peut toujours trouver, pour tout nombre $I$, un temps $T$ tel que, pour tout
 temps ult\'erieur, on trouvera une condition initiale aussi proche de la
 condition initiale choisie qui, apr\`es \'evolution, se trouvera s\'epar\'ee
 d'au moins
 $I$ de la situation finale. 
 En terme math\'ematiques cela s'exprime ainsi:
 \begin{equation}\label{sci}\nonumber
 \exists I\in\mathbb R^+ , \forall\epsilon>0,\  \exists t=t(I,\epsilon)\ \mbox{ tel que }\ \exists x,
 |x-y|\leq\epsilon,\ |\Phi^t(x)-\Phi^t(y)|\geq I,
 \end{equation}
 o\`u $\Phi^t(x)$ est le point \'evolu\'e \`a partir de $x$ au temps $t$.
 
 C'est \`a dire que l'on peut toujours trouver deux conditions initiales
 aussi proches que l'on veut qui se retrouvent s\'epar\'ees d'une distance donn\'ee apr\`es avoir
 suffisamment attendu.
 
 Une autre fa\c con de dire la m\^eme chose consiste \`a remarquer que la \sci\ implique que, si l'on
 consid\`ere pour simplifier un point fixe $y$ invariant par la dynamique (ou un point p\'eriodique,
 cela revient au m\^eme), il existe un ensemble  $\Lambda_y$ de points $x$ qui tendent vers $y$ par \'evolution \`a
 temps n\'egatif. On d\'efinit ainsi:
 \begin{equation}\label{inv}\nonumber
\Lambda_y:=\{x/
|\Phi^{-t}(x)-y|\to 0\ \mbox{ quand }\  t\to\infty\}.
\end{equation}
 La th\'eorie des syst\`emes dynamiques chaotiques permet de montrer en
 g\'en\'eral que $\Lambda_y$, d\'efinie de fa\c con ensembliste, comme classe
 d'\'equivalence, est en fait un ensemble particulier et ``gentil": c'est une vari\'et\'e
 diff\'erentiable. Dire que $\Phi^{-t}(x)\to y  \mbox{ quand }\  t\to\infty$ signifie
 que $\Phi^{-t}(x)$ a une limite,  donc que $\Phi^{-\infty}(x)$ existe pour
 tout $x\in\Lambda_y$ et que cette limite est $y,\ \forall x\in\Lambda_y$.
 
 Cela semble sugg\'erer que $\Phi^{-\infty}$ existe comme flot sur l'espace de
 phase, mais il n'en est rien. En effet, bien que le passage au temps infini soit
 tout \`a fait crucial pour les syst\`emes dynamiques (ergodicit\'e, \sci,
 syst\`emes int\'egrables), le flot \`a temps infini n'existe pas. Seules
 existent des traces ``faibles". Par exemple l'ergodicit\'e s'exprime ainsi:
 \[
 \lim_{T\to+\infty}\frac 1 {2T}\int_{-T}^Tf\circ\Phi^t=\int Fd\mu.
 \]
 On int\`egre d'abord, puis on prend la limite.
 
 Cependant, si l'on garde l'id\'ee d'un flot ``infini", on peut formellement
 ``inverser" la phrase $\Phi^{-t}(x)\to y  \mbox{ quand }\  t\to\infty$. 
 Et l'on obtient  le sch\'ema suivant:
 \vskip 0.5cm 
 \[
\Phi^{-t}(x)\to y,\ \forall x\in \Lambda_y \mbox { quand $t\to\infty$ }
\]
\centerline{$\Updownarrow$}

\[
\Phi^{-\infty}(x)=y,\ \forall x\in \Lambda_y 
\]

\centerline{$\Updownarrow$}

\[
\Phi^{+\infty}(y)=x,\ \forall x\in \Lambda_y 
\]
\centerline{$\Updownarrow$}
\vskip 0.3cm
\centerline{\ip$\ \sim\ $\id}
\vskip 1cm

Stricto sensu  cette suite d'\'equivalence n'a pas de sens, puisque
$\Phi^{\pm\infty}$ n'existe pas. Elle r\'esume cependant tr\`es bien le sens
profond de la \sci\ : la \mc, dans ses aspects chaotique, est impr\'etictible. Et
cet \ip\ fr\^ole l'\id\ dans le passage, non ``autoris\'e", au temps infini.
 
Y-a-t-il moyen de donner un sens \`a tout cela? La \mc\ est-elle, au fond,
ind\'eterminisme?

Nous allons voir bient\^ot, et ce sera le r\'esultat principal de cet article,
 que la \mc, vue comme venant de la \mq\ (\cad\ la ``vraie" situation classique),
  peut se donner elle-m\^eme une telle ``autorisation'' concernant ce ph\'enom\`ene.

 \section{L'\'evolution quantique \`a temps long}
 
 La \mq\ est lin\'eaire, c'est bien l\`a son moindre d\'efaut. On trouve parfois
 cette phrase dans la bouche de ceux qui la trouvent fade vis-\`a-vis de la \mc\ 
 chaotique. La lin\'earit\'e implique, en particulier et dans le cas de spectres
 discrets, une \'evolution quasi-p\'eriodique. Loin de toute ``chaoticit\'e" et
 autre al\'eatoire. 
 
 Or une telle vision est \'eronn\'ee pour (au moins) deux raisons : tout d'abord
 r\'eduire la \mq\ \`a une th\'eorie lin\'eaire, c'est oublier la mesure (qui ne
 l'est pas). De plus l'argument invoquant 
 la quasi-p\'eriodicit\'e ne vaut que pour un ensemble finie de fr\'equences. Ce
 qui n'est pas g\'en\'eralement le cas, et ce qui, de toute fa\c con, dispara\^\i t
 \`a la \lsc, puisque le nombre de valeurs propres par intervalle spectral fini
 augmente quand $\hbar\to 0$.
 
 Ces deux faits sont \`a la base de la discussion ci-dessous.
 
 Nous allons donc consid\'erer des situations dans lesquelles l'\'evolution sera 
 donn\'ee \`a grand temps, typiquement:
 \[
 t\sim T(\hbar),
 \]
 avec $T(\hbar)\to\infty$ quand $\hbar\to 0$, et prendre ensuite la limite
 $\hbar\to 0$.
 
 On restera donc ``quasi-p\'eriodique" pour chaque valeur de $\hbar$, mais on
 perdra cette propri\'et\'e \`a la \lsc.
 
 Remarquons que les situations concern\'ees font appel \`a des hamiltoniens
 ``confinants", \cad\ qu'il ne s'ag\^it absolument pas de th\'eorie de la
 diffusion, pour laquelle le potentiel s'\'evanouit \`a l'infini.
 
  Nous allons
 de plus nous concentrer sur la situation \'el\'ementaire o\`u l'hamiltonien a
 la forme:

\begin{equation}\label{qp1}\nonumber 
 H:=-\frac{i\hbar}2(x\frac d {dx}+\frac d {dx}x)
\end{equation}

Le flot quantique est alors obtenu par une simple dilatation :
\begin{equation}\label{nonumber}
i\hbar\partial_t\psi=H\psi \Longleftrightarrow \psi^t(x)=e^{-t/2}\psi^0(e^{-t}x).
\end{equation}

Il se trouve qu'un tel mod\`ele est symptomatique de la situation g\'en\'erale.
Il refl\`ete parfaitement l'aspect hyperbolique des situations de chaos
classique, et son extension au cas de trajectoires p\'eriodiques \`a dimension
quelconque est facile.

Nous discuterons  ce point dans la derni\`ere section de cet article,
mais voyons tout d'abord l'incidence de la formule (\ref{nonumber}) sur la \lsc.
 \section{Le passage \`a la limite et la confrontation entre \id\ quantique et \ip\ classique}
Si l'on prend pour condition initiale un \'etat coh\'erent \`a l'origine, \cad\ 
\[
\psi^0(x)=(\pi\hbar)^{-1/4}e^{-\frac{x^2}{2\hbar}},
\] on trouve facilement que :
\[
\psi^t(x)=(\pi\hbar)^{-1/4}e^{-t/2}e^{-\frac{e^{-2t}x^2}{2\hbar}},
\]
\cad\ que l'\'etat commence \`a se d\'elocaliser.

Lorsque $t\sim\frac 1 2 log(\frac 1 \hbar)$  l'\'etat est compl\`etement
decolcalis\'e:
\[
\psi^{\frac 1 2 log(\frac 1 \hbar)}(x)=\pi^{-1/4}e^{-\frac{x^2}2}.
\]
De plus lorsque
 $t\sim log(\frac 1 \hbar)$ on obtient :
\[
\psi^{log(\frac 1 \hbar)}(x)=(\frac\hbar\pi)^{1/4}e^{-\hbar\frac{x^2}2},
\]
\cad, quand $\hbar\to 0$,
\[
\psi^{log(\frac 1 \hbar)}(x)\sim(\frac\hbar\pi)^{1/4}\ .
\]
L'\'etat est uniform\'ement d\'elocalis\'e sur la droite r\'eelle,
\cad\ sur la vari\'et\'e instable associ\'ee au point fixe qui est \`a l'origine.

Voyons maintenant comment ``interpr\'eter" ce r\'esultat \`a la lumi\`ere de
l'interpr\'etation de la fonction d'onde. Si l'on effectue une mesure de
l'observable position, le r\'esultat sera $x$ avec la probabilit\'e 
$\vert\psi(x)\vert^2$. 

Mais maintenant la loi de probabilit\'e donn\'ee par $\vert\psi(x)\vert^2$ est
totalement d\'elocalis\'ee sur la droite r\'eelle (i.e. vari\'et\'e instable).

On voit donc que, en terme de mesure quantique, on peut ``affirmer" :

\[
\Phi^{+\infty}(0)=x,\ \forall x\in \Lambda_0:=\mbox{ droite r\'eelle. } 
\]
\vskip 0.3cm
\centerline{\bf On a donc autoris\'e la cha\^\i ne interdite de la section \ref{sci}.}
\vskip 0.3cm

Au fond qu'avons-nous fait?

Nous avons appliqu\'e les stricts postulats de la \mq\ : mesure et \'evolution.

Au d\'ebut l'\'etat est localis\'e en un point (\et) et la mesure, dans la \lsc,
donne ce point comme r\'esultat, avec probabilit\'e $1$.

Puis nous avons laiss\'e cet \'etat \'evoluer jusqu'\`a un temps de l'ordre de
$\log{\frac 1 \hbar}$.

Puis nous avons effectu\'e la mesure de la position \`a nouveau.

Et l\`a la probabilit\'e est devenue uniform\'ement repartie sur la
vari\'et\'e instable, en contradiction apparente avec le paradigme classique,
mais en accord avec l'esprit de la \sci.

\section{Invariants et stabilit\'e \`a l'int\'erieur du chaos}
Nous venons de voir que, pour l'\'evolution quantique \`a grand temps, la fonction d'onde peut, alors
qu'\`a l'origine elle \'etait concentr\'ee en un point, se d\'elocaliser sur la vari\'et\'e instable
issue de celui-ci. La g\'en\'eralisation de ce ph\'enom\`ene concerne la notion de vari\'et\'e instable
associ\'ee \`a une trajectoire quelconque \cite{DE}. On d\'efinit donc maintenant :
\[
\Lambda_y^\pm:=\{x/
|\Phi^{\pm t}(x)-\Phi^{\pm t}y|\to 0\ \mbox{quand}\  t\to\infty\}.
\]
Les vari\'et\'es stables et instables sont covariantes par le flot dans le sens que :
\[
\Phi^s(\Lambda^\pm_y)=\Lambda^\pm_{\Phi^{-s}y}.
\]
En particulier $\cup_{s\in\mathbb R}\Phi^s(\Lambda^\pm_y)$ est invariant par le flot.

On voit donc que la \mq\ a la propri\'et\'e, lors de l'\'evolution \`a grand temps et la \lsc, de
d\'elocaliser des \et, initialement localis\'es en un point (\'el\'ement d'un espace absolu), 
en des \'etats localis\'e sur les
\textit{in- ou co-variants} de la dynamique classique. Ces nouveaux objets, les invariants de la
dynamique, sont en quelque sorte les \'el\'ements d'un nouvel espace absolu, l'espace des invariants de
la dynamique. 

Ces objets ne sont pas nouveaux, ce sont m\^eme les outils fondamentaux pour \'etudier la dynamique
chaotique,  donc \`a temps long, mais ils gardent dans la \mc\ un statut d'outil, palliatif au fait que
l'on ne sait plus calculer, ou m\^eme appr\'ehender, pour de telles \'echelles de temps, la notion de trajectoire.

La \mq, ayant d\`es le d\'epart abandonn\'e dans son formalisme la notion de trajectoire au sens de la
g\'eom\'etrie du point, leur donne, \`a ces invariants, un vrai statut ontologique.

Mais il n'y a pas de raison que cela s'arr\^ete l\`a. Puisque la \mq\ a transform\'e le point en vari\'et\'e
instable, elle doit, pour des \'echelles de temps encore sup\'erieures, transformer \`a nouveau ces nouveaux
objets en quelque chose d'autre. C'est ce qui se passe dans l'exemple du ``$8$" \cite{p1}.
L'exemple des dilatations faisait en effet partir tout \`a l'infini (temporel et spatial). Mais tout en
gardant l'infini temporel, il est possible de compactifier l'aspect spatial en consid\'erant l'exemple
suivant.
Consid\'erons maintenant un hamiltonien de la forme :
\[
h(q,p)=p^2+q^2(q^2-1).
\]
Cela correspond \`a une particule dans un potentiel de la forme d'un double puit.
Les deux vari\'et\'es  instables sont les deux branches d'un ``$8$" inclin\'e :
les trois invariants de la dynamique sont les deux branches du $8$ et le point fixe.

On peut dans ce cas calculer le flot, dans la \lsc\  \`a temps multiple de 
$t=\log{\frac1 \hbar}$. On trouve ainsi un
mouvement
entre les deux branches du $8$ et le point fixe, au milieu. Cette nouvelle dynamique est une dynamique sur les
invariants de la dynamique classique, comme la dynamique pr\'ec\'edente l'\'etait sur les points de l'espace
absolu.

\section{Conclusion et lien avec d'autres disciplines}
Nous avons essay\'e dans cet article de montrer comment les notions,  en principe tellement diff\'erentes, d'ind\'eterminisme quantique et
d'impr\'edictiboilt\'e classique pouvaient se superposer dans la limite des grands temps d'\'evolution
coupl\'ee \`a la \lsc. Il est apparu que l'ind\'eterminisme quantique, et son aspect probabiliste 
donn\'e par l'interpr\'etation statistique de la fonction d'onde, est non seulement en accord avec la \mc, mais qu'il donne
une justification \`a la notion de flot \`a temps infini en \mc\ : la sensibilit\'e aux conditions initiales peut \^etre en effet 
vue, dans la limite o\`u le temps d'\'evolution diverge, soit comme ind\'eterministe (un point donnant lieu \`a 
une vari\'et\'e instable), soit comme
flot g\'en\'eralis\'e transformant un point en un ensemble de dimension sup\'erieure. 
La \mq\ unifie les deux gr\^ace \`a l'interpr\'etation probabiliste
 coupl\'ee \`a la d\'elocalisation de la fonction d'onde.
 \vskip 0.5cm
 Il appara\^\i t ainsi, me semble-t-il, une notion ontologique donn\'ee \`a la notion d'invariant de  la dynamique. Les vari\'et\'es 
 instables, invariants de la dynamique, deviennent ainsi un objet \'epais, important : le lieu des possibles lorsque l'on effectue
 une mesure quantique, le support de la loi de probabilit\'e.

Ce couplage, qui d\'etruit la notion d'espace absolu habituelle pour la remplacer par celui des
invariants de la th\'eorie classique, fait appara\^itre le r\^ole important jou\'e par ces dits invariants.
\vskip 0.5cm
L'\'etude des invariants d'une th\'eorie est un grand standard de la science moderne. On le retrouve en 
permanence en math\'ematique et physique, mais aussi en biologie etc.
Remarquons pour finir que ce jeu qui consiste \`a embo\^\i ter une nouvelle dynamique sur l'espace des
invariants de la pr\'ec\'edente sugg\`ere que le caract\`ere principal de la \mq, qui est la stabilit\'e,
en principe incompatible avec l'hypoth\`ese chaotique de la \mc, pourrait bien \^etre finalement
expliqu\'ee, dans des cas simples, par une dynamique stable sur les invariants de la dynamique instable. C'est
en tous cas ce que nous a montr\'e l'exemple du ``8".

\end{document}